\documentclass[lettersize,journal]{IEEEtran}
\usepackage{epsfig,setspace,amsmath,epsf,amssymb,bm,theorem,cite,graphicx, epstopdf,algorithm,float,color,mathtools, authblk}
\usepackage[table,xcdraw]{xcolor}
\usepackage{subcaption}
\usepackage{algpseudocode}

\usepackage{multirow}
\usepackage{arydshln}

\usepackage[short,c2]{optidef}

\newtheorem{theorem}{Theorem}

\newcommand{\e}{{\mathbb{E}}}

\IEEEoverridecommandlockouts
\allowdisplaybreaks

\begin{document}

\title{Scheduling Policies in a Multi-Source Status Update System with Dedicated and Shared Servers}
\author[1]{Sahan Liyanaarachchi}
\author[1]{Sennur Ulukus}
\author[2]{Nail Akar}

\affil[1]{\normalsize University of Maryland, College Park, MD, USA}
\affil[2]{\normalsize Bilkent University, Ankara, T\"{u}rkiye}
\maketitle

\begin{abstract}
Use of multi-path network topologies has become a prominent technique to assert timeliness in terms of age of information (AoI) and to improve resilience to link disruptions in communication systems. However, establishing multiple dedicated communication links among network nodes is a costly endeavor. Therefore, quite often, these secondary communication links are shared among multiple entities. Moreover, these multi-path networks come with the added challenge of out-of-order transmissions. In this paper, we study  an amalgamation of the above two aspects, i.e., multi-path transmissions and link sharing. In contrast to the existing literature where the main focus has been scheduling multiple sources on a single shared server, we delve into the realm where each source sharing the shared server is also supplemented with its dedicated server so as to improve its timeliness. In this multi-path link sharing setting with generate-at-will transmissions, we first present the optimal probabilistic scheduler, and then propose several heuristic-based cyclic scheduling algorithms for the shared server, to minimize the weighted average age of information of the sources.
\end{abstract}

\section{Introduction}
Timeliness has become an indispensable feature that needs to be integrated into communication systems spanning from internet of things (IoT)  applications \cite{iot_aoi_2019} to cislunar communications \cite{yuan_towards, sahan-cislunar}. Vehicular networks used in autonomous driving, remote surgery systems, uncrewed lunar landing missions are a few avenues where timely communication is critical \cite{yuan_towards, sahan-cislunar, AoI_self_drive, tactile_internet}. Therefore, the design of network infrastructure and the development of sampling and scheduling policies to improve the timeliness of communication, has been a broadly pursued research direction in the recent literature.    

Age of information (AoI) has become a prominent metric of interest to quantify timeliness in communication systems \cite{RoyYates__AgeOfInfo_Survey, Kosta_AoI_2017,Sun_AoI_2019}. The AoI or simply the instantaneous age, denoted by $\Delta(t)$, measures the time that has elapsed from the time of generation of the latest received update, and is given by $\Delta(t)=t-g(t)$, where $g(t)$ is the generation time of the latest received update. To model the AoI process, communication channels (or links) are often modeled as queuing systems where the service time of the server corresponds to the delay experienced in the link. 

A majority of the existing literature on timeliness of communication revolves around status update systems with a single server, where the primary focus has been to devise policies for sampling the stochastic process associated with status generation so as to minimize the time averaged AoI \cite{update_wait}. In addition, for the case of multiple sources, significant attention has been given to the development of scheduling policies for source transmissions in this single-server setting \cite{rtsms2019, tme2022, tmef2023}. As an outcome of these efforts, a wide spectrum of age-dependent scheduling policies, such as max-weight, Whittle-index, maximum-age-first (MAF) and maximum-age-difference-drop (MAD) \cite{age_aware1, age_aware2, maf1, maf2, maf3, mad}, as well as age-agnostic scheduling policies, such as cyclic and probabilistic scheduling schemes \cite{gau23, akar_etal_infocom24, NOTS2024} have emerged in this setting. Moreover, some of these efforts have recently shifted towards the analysis and optimization of status update systems with path diversity, i.e., systems with sources receiving service from multiple servers. Path diversity has become a simple but effective design technique to enhance the timeliness of communication \cite{path_diversity}. As an example, the work in \cite{Pappas_dual_server} obtains an expression for the average AoI of a single-source dual-server system, where it is shown that the average AoI improves by $37.5\%$ when the service rates of the two servers are identical.

\begin{figure}
    \centering
    \includegraphics[width=0.8\columnwidth]{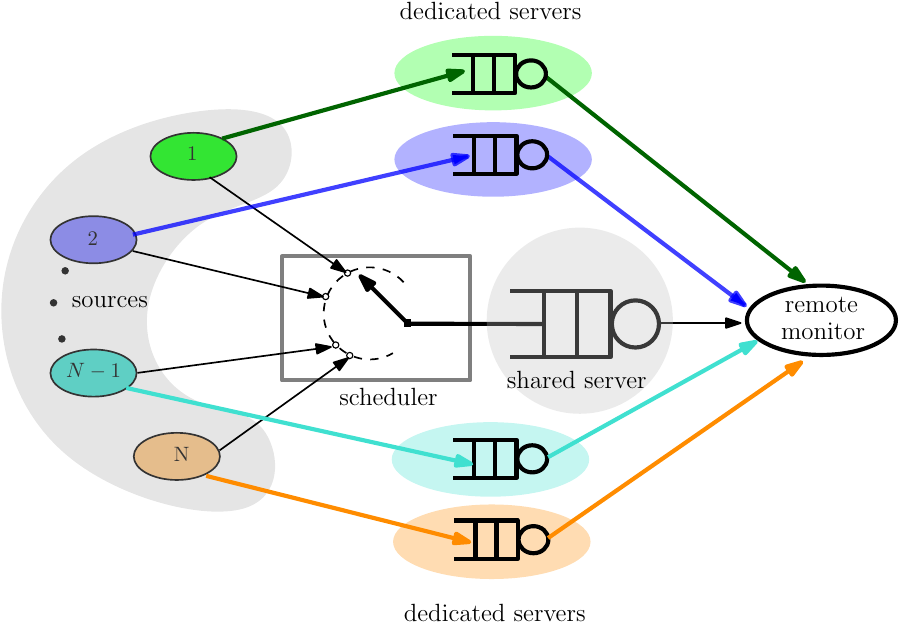}
    \caption{Shared server status update system with $N$ sources.}
    \label{fig:sys_model}
\end{figure}

Despite their efficacy, multi-server architectures suffer from a phenomenon known as \emph{out-of-order transmissions} since the packets may not have to be received in the same order they were generated due to the availability of multiple paths for a single information source. Even though various techniques such as stochastic hybrid systems (SHS) \cite{rtsms2019} and absorbing Markov chain (AMC) \cite{nailAMC2023} formulations have emerged  to facilitate the AoI analysis of such queuing systems, the analysis can still be arduous in the case of multiple sources and out-of-order transmissions. Therefore, most recent works have been limited to either the AoI analysis of single-source multi-server systems, or the construction of scheduling policies for a multi-source single-server system.

In this work, different from the approaches developed so far, we study a status update system given in Fig.~\ref{fig:sys_model} that brings together the essence of  single-source multi-server and multi-source single-server systems, while accounting for out-of-order transmissions. In particular, we envision a system where multiple sources are scheduled through a single shared server, but in addition, each source has its own dedicated server with its unique service rate which may be different than the service rates of other sources.

Systems using both dedicated and shared servers, i.e., hybrid status update systems, arise naturally in various application scenarios including wireless relay networks \cite{relay_net} and  mobile edge computing \cite{Melih_mean_field}. In a relay network, having multiple links between nodes can notably improve timeliness. However, establishing several dedicated links between individual nodes can be costly. Therefore, some links need to be shared among multiple entities to minimize these costs. Thus, path diversity is achieved through link sharing in most conventional relay networks. The work in  \cite{yuan_towards} studies one such system in the context of cislunar communications, where the authors consider the sharing of two satellite relays among multiple lunar probes/sensors to improve the timeliness of communication with the lunar far side. Moreover, in such cislunar scenarios, due to long propagation delays, maintaining an exact replica of the AoI process can be an arduous task as it requires multiple round trip communications across vast distances and this can significantly impede system performance. As highlighted in \cite{igor_2025}, even the slightest communication delay can adversely affect the performance of age-aware policies. For such applications, age-agnostic scheduling is deemed to be a promising alternative. In the realm of edge computing, edge servers are used to reduce latency by processing the data at close proximity and hence have become a vital component in industrial IoT applications \cite{afzal_iiot}. In these edge computing applications, each dedicated server can be viewed as an end device with its own local compute power, and the shared server represents the mobile edge server that is shared among the end devices. To improve the overall latency, each device may either locally process the data, or offload some of the processing to the edge server. In here, the local processing power of each end device is congruous to dedicated servers having their unique service rates, while edge computing power is equally available to each device which is coherent with the fact that each source experiences the same service rate at the shared server. The works in \cite{MEC_1,MEC_2,MEC_Yates} are a few other avenues on computation offloading in mobile edge computing systems, which align with the envisioned system given in Fig.~\ref{fig:sys_model}.

We study the hybrid status update system given in Fig.~\ref{fig:sys_model} with \emph{generate-at-will} (GAW) transmissions, and we present several \emph{age-agnostic} scheduling policies for source transmissions on the shared channel, for the case of exponentially distributed service times for all server types. To summarize our contributions:
\begin{itemize}
    \item We rigorously analyze the AoI of the hybrid status update system  under the GAW model, and we provide expressions for the average AoI of each source under a given age-agnostic \emph{cyclic} or \emph{probabilistic} scheduling scheme. For this purpose, we depart from the conventional graph-based methods to compute the AoI, and instead use the AMC formulation to tackle the complications arising from out-of-order transmissions.
    \item We pose the construction of the optimal probabilistic scheduler as a convex optimization problem by using the obtained closed-form expressions under probabilistic scheduling. Moreover, we provide a water-filling algorithm to compute the optimal scheduling probabilities.
    \item We present several heuristic-based cyclic schedulers which have superior AoI performance compared to the optimum probabilistic scheduler.
\end{itemize}

The remainder of the paper is organized as follows. Section~\ref{sec:rel_work} provides a summary of the related work. Section~\ref{sec:sys_mod} outlines the system model. Section~\ref{sec:pol} describes the scheduling policies used in this work. Section~\ref{sec:aoi} gives the AMC formulation for the AoI analysis of the considered scheduling policies. Sections~\ref{sec:opt_PS} and \ref{sec:cyc} study the construction of the optimal probabilistic scheduler and the heuristic-based cyclic schedulers, respectively. In Section~\ref{sec:results}, we provide the numerical validation of our schemes, and finally, we conclude and discuss future work items in Section~\ref{sec:con}. 

\section{Related Work}\label{sec:rel_work}

Most of the prior works on path diversity and AoI have concentrated around finding the average AoI of single-source multi-server status update systems. One of the earliest works in this domain roots back to \cite{Yates_tadam_queues} which uses an SHS formulation to derive the expression for the average AoI of queues in tandem and later this analysis was expanded to queues in parallel in \cite{Yates_parallel_queues}. {Both of these works revolved around the \emph{random arrival} (RA) model where status packet generation is governed by a Poisson arrival process.

The work in \cite{path_diversity} analyzes the effect of path diversity on status age where the authors first model a network with  ample resources as a $\text{M/M/}\infty$ queuing system while accounting for out-of-order transmissions. Then, this analysis is extended for a dual server system which is modeled as a \emph{first-come-first-serve} (FCFS) $\text{M/M/2}$ queuing system where they derive expressions for the mean AoI by approximating the distributions of the inter-arrival times of informative updates. On a similar vein of research, in \cite{Sun_multipath}, the authors study a multi-server queuing system where incoming packets are enqueued before transmission, and they show that under the considered system model, a \emph{preemptive last-generated-first-served} (LGFS) policy optimizes both AoI and throughput of the system. Both these works operate under the RA model for status update packet generation. In the recent past, there have been a few works which study the problem of scheduling to minimize the AoI of multi-source multi-server systems. However, these works often rely on the assumption that only one server will cater to a single source at any given time so as to avoid the complications arising from out-of-order transmissions \cite{Sun_markov_gaus}.

SHS method was used in \cite{Pappas_dual_server} to obtain the average AoI of dual queue/server status update system under the GAW model introduced in \cite{lts2015} for status generation. In \cite{nailAMC2023}, the AMC method was introduced as an alternative to SHS and it was used to derive the exact distribution of the steady-state random variable associated with the AoI process of a single-server queue under a GAW model,  where the higher order moments of the AoI process emerged as a natural outcome while employing this method. This method was later utilized in \cite{nailDualServer2024} to develop a freeze and preemption policy to further improve the AoI of the dual-server status update system introduced in \cite{Pappas_dual_server}. Finally, the effect of multiple parallel transmissions on the timeliness of status update systems was studied in \cite{Pappas_parallel_queues} by employing  the SHS method. All these works demonstrate the  value of path diversity in networks for timely communication, resulting in substantial reduction in the average AoI.

The closest to our work is \cite{Melih_mean_field}, where the authors study a  system of multiple sources where each source probabilistically opts to either process the data locally or through a shared edge server. This scenario was modeled as a push-based queuing system with multiple parallel servers and  a single shared server under an RA  status generation process. Moreover, the shared server is assumed to follow a \emph{last-come-first-serve} (LCFS) with \emph{preemption} policy which allows the use of the SHS method to find the average AoI and employ a mean field game model to optimize for the processing power and the selection probabilities of individual sources in the large population regime. This work does not encapsulate the essence of a scheduling problem but rather that of a game theory problem, where each source tries to maximize a local objective while accounting for the interference from other sources. In contrast to \cite{Melih_mean_field}, in this work, we study a pull-based status update system with non-preemptive servers under a GAW model where we employ an AMC formulation to derive expressions for the AoI of the system explicitly focusing on the framework of age-agnostic scheduling.

\section{System model} \label{sec:sys_mod}
Consider the multi-source dual-server (one dedicated server and one shared server, for each source) status update system shown in Fig.~\ref{fig:sys_model}, where each source has a dedicated channel to the remote monitor along with access to a single channel which is shared among all the sources based on a suitable scheduling policy. All of the channels are modeled with exponentially distributed service times. We assume a GAW model for the status generation process where each server immediately sends a pull request to one of the sources once it has finished transmitting (serving) the previous source packet. Thus, in this system, we have work-conserving servers that never idle.

Let $S_n$ for $n \in \{1,2,\dots,N\}$ denote the direct channel (dedicated server) between source-$n$ and the remote monitor with an exponentially distributed service time with parameter $\mu_n$. Let $S$ denote the shared channel (shared server) among the sources whose exponentially distributed service time for any of the sources has the service rate $\mu$. We denote by $\Delta_n(t)$ the associated AoI process of source-$n$ updates, which is a random process which linearly increases with time until the remote monitor receives an update from source-$n$ with a fresher timestamp upon which the process drops to the service delay of the freshest source-$n$ update. Since each source has two possible pathways for transmission, the updates that arrive at the monitor, can occasionally be out-of-order. In the event that we receive an older timestamp due to out-of-order transmissions, that particular update will simply be discarded by the monitor without modifying the age of that particular source.

We define $\Delta = \sum_{n=1}^N w_n\Delta_n$ as the weighted AoI where $\sum_{n=1}^N w_n = 1$ and $w_n$ is the source-specific weight assigned to source-$n$. Here, $\Delta_n(t)$ is the AoI process for source-$n$ maintained at the remote monitor and $\Delta_n$ is the steady-state random variable associated with the process $\Delta_n(t)$. The goal of this work is to devise scheduling policies for source transmissions across the shared server so as to minimize the expected weighted AoI of the status update system. 

\section{Scheduling Policies}\label{sec:pol}
Source transmissions need to be scheduled appropriately on the shared server to minimize the expected weighted AoI. Since we consider a GAW model with work-conserving servers, the scheduler must decide which source it must schedule once the current transmission through the shared server has finished. For each server/channel, we assume that the remote monitor only provides feedback for their respective channel transmissions and nothing more. Due to the lack of perfect communication between the scheduler and the dedicated servers, the scheduler does not know the ages perfectly which renders any estimates of the AoI processes made at the scheduler side inaccurate, and makes the use of age-dependent scheduling unsuitable. Therefore, in this paper, we consider only age-agnostic policies due to their simplicity, and also the lack of a need in such policies to maintain an exact replica of the AoI processes at the scheduler. Moreover, due to imperfect communication, the server lacks the necessary information to determine whether the currently transmitted data packet is obsolete or not. Hence, the packets can be discarded only once they reach the remote monitor.

Next, we describe the two age-agnostic scheduling schemes studied in this paper.

\subsection{Probabilistic Scheduler}
In the probabilistic scheduler (PS), once the shared server becomes free, the next source transmission is selected based on a probability mass function (pmf). Let $\{p_1,p_2,\dots,p_N\}$ denote this pmf. In particular, source-$n$ is scheduled for transmission with probability $p_n$ once the current transmission is over.

\subsection{Cyclic Scheduler}
In the cyclic scheduler (CS), source transmissions are scheduled based on a fixed finite pattern which repeats itself. For example, let $N=4$ and also let $C=[1,2,2,3]$ be the cyclic pattern and $C_n$ be a sample obtained from source-$n$. Then, the source transmissions on the shared server will be $C_1, C_2, C_2, C_3, C_1, C_2, C_2, C_3, \dots$. Note that, even though we have four sources, based on the pattern $C$, we will never allocate a scheduling instance for the fourth source. This is to highlight the fact that any pattern which schedules at least one source schedule instance is a feasible cyclic schedule. In other words, the shared server need not cater to all sources.

\begin{figure}[t]
    \centering
    \includegraphics[width=0.75\columnwidth]{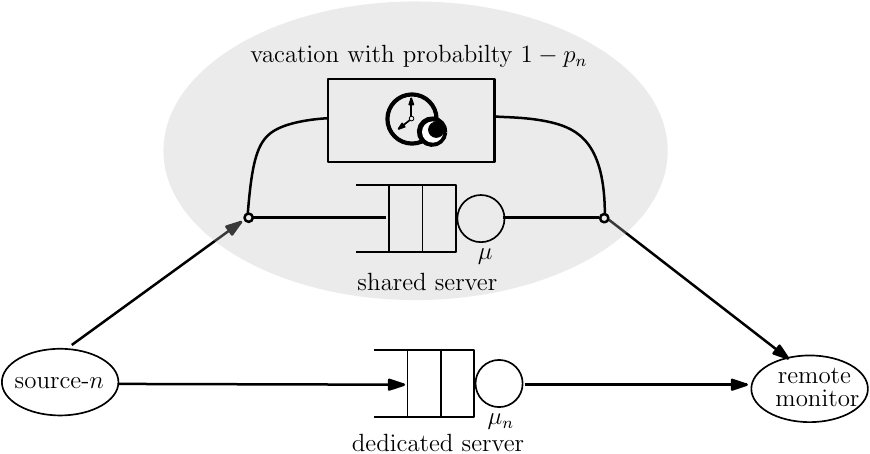}
    \caption{Dual-server sub-problem for PS.}
    \label{fig:2_server_sys}
\end{figure}

\section{AoI Analysis}\label{sec:aoi}
The AoI of source-$n$ depends only on the transmissions emanating from its dedicated server and the shared server. When the shared server is occupied by another source transmission, from the perspective of source-$n$, it is as if the shared server is taking a vacation with rate $\mu$. Therefore, one can reduce the $N$-source, $(N+1)$-server problem into $N$ single-source dual-server problems, where the shared server takes a vacation based on either some probability (for PS) or according to a cyclic schedule (for CS) from the perspective of source-$n$. Fig.~\ref{fig:2_server_sys} illustrates the dual-server sub-problem for the PS.

Finding the average AoI of the dual-server problem is not a straighforward task. Due to the out-of-order arrival of packets at the remote monitor, the traditional graphical area based computation is not feasible in this scenario. Therefore, we employ the absorbing Markov chain (AMC) formulation, which was introduced in \cite{nailAMC2023}, for exact analytical modeling of AoI.

The key idea of the AMC method is to model the sample path followed by a newly joining packet into the system until it is successfully received (without being obsolete) by the remote monitor. Note that a single AoI cycle begins upon the successful reception of a packet and will continue to linearly increase until the next successful reception. Let $P$ be a packet that just entered the system. At this time instance, we will initiate our AMC and we let it evolve until $P$ is successfully received. The time at which $P$ becomes successful will correspond to the beginning of a typical AoI cycle (see Fig.~\ref{fig:aoi_cycle}). In this case, we let the AMC further evolve until the next successful reception of a packet and consider this to be one absorbing state of our AMC. This absorbing state corresponds to a successful reception of a new packet following the successfully-received packet $P$, and therefore, is termed as the successful absorbing state. In the event that $P$ becomes obsolete because a packet with a later timestamp has arrived at the remote server before $P$, then we consider that the AMC gets absorbed into another absorbing state termed as the unsuccessful absorbing state. 

\begin{figure}[t]
    \centering
    \includegraphics[width=0.7\columnwidth]{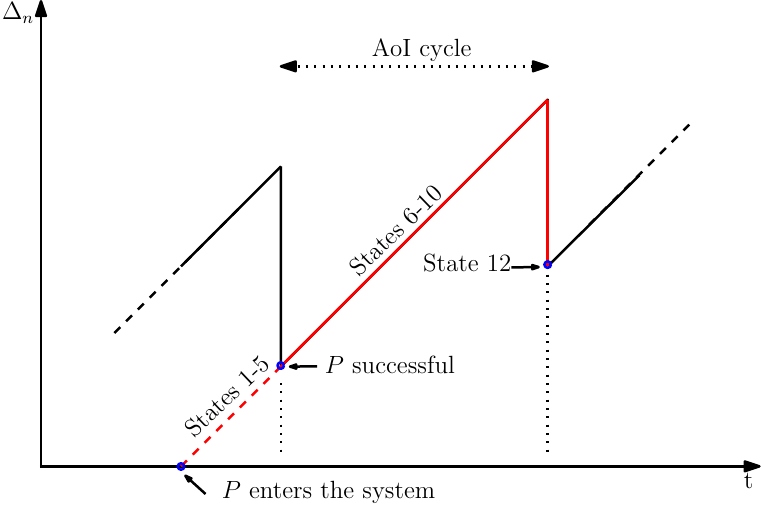}
    \caption{Sample path of the AoI of source-$n$ with the subsequent states of the PS AMC.}
    \label{fig:aoi_cycle}
\end{figure}

To compute the average AoI using the AMC method, we require the generator matrix of the AMC and the initial probability vector of the transient states of the AMC. Therefore, the AoI analysis using an AMC involves three main steps for a given source-$n$:
\begin{enumerate}
   \item construction of the AMC, 
   \item construction of a recurrent Markov chain (RMC) to compute the initial probabilities of the AMC,
   \item computation of the mean of $\Delta_n$.  
\end{enumerate}
This procedure needs to be repeated for all sources.
 
Let $\bm{Q}$ be the generator matrix of the AMC obtained in the first step above, which is defined as follows,
\begin{align}
     \bm{Q} = 
     \left[
     \begin{array}{c:c}
         \bm{U} & \bm{V} \\
         \hdashline
         \bm{0}& \bm{0} 
     \end{array}
   \right],
\end{align}
where $\bm{U}$ is the sub-generator matrix governing transitions among the transient states, $\bm{V}$ is the sub-generator matrix representing the transitions from the transient states to the two absorbing states, and $\bm{0}$ stands for a matrix of zeros of appropriate size. Let $\bm\sigma$ be the initial probability row vector of the transient states. Then, it is shown in \cite{nailDualServer2024} that the average AoI for source-$n$ can be found as follows,
\begin{align}\label{eqn:aoi_form}
    \e[\Delta_n] = - \frac{\bm{\sigma}\bm{U}^{-2}\bm{\theta}}{\bm{\sigma}\bm{U}^{-1}\bm{\theta}},
\end{align}
where $\bm{\theta}$, termed as the transient vector, is a binary column vector of the same size as $\bm{\sigma}$ with ones in the entries corresponding to the transient states during which the packet $P$ (which initiated the AMC in the first place) had already been successfully received by the remote monitor. On the other hand,  the components of $\bm{\theta}$ corresponding to  the states during which the original packet $P$ is still in the system, are zero. The distribution and higher-order moments of $\Delta_n$ can also be obtained using the method of \cite{nailAMC2023} but our focus in this paper is only on the mean AoI for source-$n$.

Now, we will present the AMC construction for the dual-server sub-problems for PS and CS.

\begin{table}[]
    \centering
    \caption{States of AMC for the PS sub-problem of source-$n$.}
    \label{tab:PS_states}
    \begin{tabular}{|c|c|}
    \hline
         State& Description  \\
         \hline
         1& $P$ on $S_n$, $S$ on vacation \\
         \hline
         2& $P$ on $S_n$, $S$ busy, $T_n\geq T_s$ \\
         \hline
         3& $P$ on $S_n$, $S$ busy, $T_n < T_s$ \\
         \hline
         4& $P$ on $S$, $S_n$ busy, $T_n\leq T_s$ \\
         \hline
         5& $P$ on $S$, $S_n$ busy, $T_n > T_s$ \\
         \hline
         6& $P_n$ on $S_n$ up to date, $S$ on vacation \\
         \hline
          7& $P_n$ on $S_n$ up to date, $P_s$ on $S$ obsolete \\
         \hline
          8& $P_n$ on $S_n$ obsolete,  $P_s$ on $S$ up to date \\
         \hline
         9& $P_n$ on $S_n$ up to date, $P_s$ on $S$ up to date \\
         \hline
         10 & $P_n$ on $S_n$ obsolete, $S$ on vacation \\
         \hline
         11 & Unsuccessful absorbing state \\
         \hline
         12 & Successful absorbing state \\
         \hline
    \end{tabular}
    
\end{table}

\subsection{Probabilistic Scheduler AoI}\label{sec:ps_aoi}
We analyze the AoI of source-$n$ when PS is employed for the dual-server sub-problem first. For source-$n$, once the shared server becomes free, it will pull a packet from source-$n$ with probability $p_n$, or will go on a vacation with probability $q_n=(1-p_n)$. Let $P$ be the packet that initiates the AMC and let $P_n$ and $P_s$ be  typical packets in server $S_n$ and server $S$, respectively. Let the timestamps of the packets in $S_n$ and $S$ be denoted by $T_n$ and $T_s$. Based on at which server a newly arriving packet $P$ joins the system, the corresponding timestamps of packets in the servers, and whether $S$ is on a vacation or not, we identify  12 states for the evolution of the AMC. These states are given in Table~\ref{tab:PS_states}. 

The transitions between the states occur when either the dedicated server becomes free and  pulls a new source-$n$ packet into the system, or when the shared server becomes free and decides either to go on a vacation or pull a new source-$n$ packet. The state transitions of the above AMC are as follows:
\begin{itemize}
    \item When in state 1, the packet $P$, which initiated the AMC, will be successfully received by the remote monitor with rate $\mu_n$. In this case, a new up-to-date (not obsolete) packet will be pulled from source-$n$ into the dedicated server, and hence, the ACM transitions to state 6 with rate $\mu$. In state 1, the shared server, which is on vacation, will pull a new packet from source-$n$ with rate $p_n\mu$. This new packet will have a later timestamp than $P$, and hence, the AMC transitions to state 3 with rate $p_n\mu$.
    \item In state 2, $P$ on $S_n$ will be successful  and  a new packet will be pulled into server $S_n$ with rate $\mu_n$. Since the packet $P$ has a later timestamp than the packet in $S$, once $P$ becomes successful, the packet in $S$ will become obsolete. Therefore, the AMC will transition to state 7 with rate $\mu_n$. In state 2, the shared server will pull a new packet with rate $p_n\mu$ or will go into a vacation with rate $q_n\mu$. If it pulls a new packet, then the packet in the shared server will have a later timestamp than the packet in server $S_n$. Therefore, the AMC will transition to state 3 with rate $p_n\mu$ and to state 1 with rate $q_n\mu$.
    \item In state 3, $P$ on $S_n$ will be successful and a new fresher packet will be pulled into server $S_n$ with rate $\mu_n$. In this case, since $P$ has an earlier timestamp than the packet in $S$, the packet in $S$ will be up-to-date even after the reception of $P$. Hence, the AMC will transition to state 9 with rate $\mu_n$. Moreover, in state 3, the packet in $S$ will be successful with rate $\mu$, which will make the packet $P$ on $S_n$ obsolete since it has an older timestamp. Since $P$ becomes obsolete, the AMC will be absorbed to state 11 with rate $\mu$.
    \item In state 4, $P$ on $S$ will be successful and a new packet will be pulled onto $S$ with rate $p_n\mu$ or $P$ on $S$ will be successful, and $S$ will go onto a vacation with rate $q_n\mu$. In either case, once $P$ becomes successful, the packet in $S_n$ which has an older timestamp than $P$ will become obsolete. Therefore, the AMC will transition to state 8 with rate $p_n\mu$ and to state 10 with rate $q_n\mu$. Additionally, when in state 4, the packet in $S_n$ will be successful with rate $\mu_n$ and in this case a new packet with a fresher timestamp will be pulled into $S_n$. Therefore, the AMC will transition to state 5 with rate $\mu_n$.
    \item In state 5,  $P$ on $S$ becomes successful and a new packet will be pulled into $S$ with rate $p_n\mu$ or $P$ becomes successful and $S$ goes onto vacation with rate $q_n\mu$. If a new packet was pulled onto $S$, since $P$ had an older timestamp than the packet in $S_n$, both the new packet and the packet in $S_n$ will be up to date and hence the ACM will transition to state 9 with rate $p_n\mu$. On the other hand, if $S$ goes into a vacation once $P$ is successful, the AMC will transition to state 6 with rate $q_n\mu$ since the packet on $S_n$ is still up-to-date. When in state 5, the packet in $S_n$ which has a fresher timestamp than $P$ will be successful with rate $\mu_n$. In this case, the packet $P$ will become obsolete and hence the AMC will be absorbed onto state 11 with rate $\mu_n$.
    \item When in states 6 to 10, the corresponding packet $P$ which initiated the AMC has been successful. In these states, once an up-to-date packet finishes its transition, the AMC will be absorbed into state 12. If an obsolete packet finishes its transition, it will be discarded and a new packet will replace the obsolete packet in the corresponding server. If $S$ is on a vacation, then a new packet will be pulled into $S$ with rate $p_n\mu$.
\end{itemize}

\begin{table}[]
    \centering
    \caption{AMC transition rates  for  PS sub-problem.}
    \label{tab:PS_rates}
    \begin{tabular}{|c|c|c|c|c|c|}
    \hline
    \multicolumn{3}{|c|}{Transition rates} & \multicolumn{3}{c|}{Transition rates}\\
    \hline
     From & To & Rate & From & To & Rate \\
    \hline
    \multirow{2}{*}{1} & 6 & $\mu_n$ & \multirow{2}{*}{6} & 12& $\mu_n$\\
                       \cline{2-3} \cline{5-6}
                       & 3 & $p_n\mu$& & 9&$p_n\mu$ \\
    \cline{1-6} 
    \multirow{3}{*}{2} & 7 & $\mu_n$ & \multirow{3}{*}{7} &12 & $\mu_n$ \\
                       \cline{2-3} \cline{5-6}
                       & 3 & $p_n\mu$ & &9 & $p_n\mu$\\
                       \cline{2-3} \cline{5-6}
                       & 1 & $q_n\mu$ & &6 &$q_n\mu$ \\
    \cline{1-6}
    \multirow{2}{*}{3} & 9 & $\mu_n$ & \multirow{2}{*}{8} & 9& $\mu_n$\\
                       \cline{2-3} \cline{5-6}
                       & 11 & $\mu$& & 12 & $\mu$ \\
    \cline{1-6}
    \multirow{3}{*}{4} & 5 & $\mu_n$ & 9 &12 &$\mu_n + \mu$ \\
                       \cline{2-6}
                       & 8 & $p_n\mu$ & \multirow{2}{*}{10}&6 &$\mu_n$\\
                       \cline{2-3} \cline{5-6}
                       & 10 & $q_n\mu$ & &8 & $p_n\mu$\\
    \cline{1-6}    
    \multirow{3}{*}{5} & 11 & $\mu_n$ & \multicolumn{3}{c|}{} \\
                       \cline{2-3}
                       & 9 & $p_n\mu$ & \multicolumn{3}{c|}{}\\
                       \cline{2-3}
                       & 6 & $q_n\mu$ & \multicolumn{3}{c|}{} \\
    \cline{1-6}    
    
    \end{tabular}
    
\end{table}

The above transition rates of the transient states are summarized in Table~\ref{tab:PS_rates}. To better understand how these state transitions relate to the age graph, we provide three example scenarios highlighting the differences between successful and unsuccessful absorptions  in Fig.~\ref{fig:AMC_examples}. In all  three scenarios, we assume that the newly joining packet $P$ enters $S_n$ (at time $t_0$) and at that time instance, $S$ is busy with an up-to-date packet. Therefore, the initial state of our AMC is state 2. The servers that finish service at specific time instances are depicted using downward facing arrows on top of the age curve. For example in all three scenarios, at time $t_1$, server $S$ finishes service and  pulls a new packet into it, or goes into vacation.

Fig.~\ref{fig:amc_success_1} illustrates a scenario where the initiated AMC at time $t_0$, ends up in an successful absorption. Here, at time $t_1$,  $S$ finishes its service at time $t_1$, and since it had an up-to-date packet in it, the age drops. In here, we consider the scenario where a new packet from source-$n$ is pulled into $S$ at time $t_1$ causing  the AMC to transition to state 3. Since the packet in $S_n$ had entered the system at a later time instance than the most recently received  packet from $S$, at time $t_n$, $S_n$ will finish its transmission, and will result in an age drop. Then, a new packet will join $S_n$ and the AMC will transition to state 9 since both the packets in $S$ and $S_n$ are still up-to-date. Subsequently, since both servers contain up-to-date packets, when either one of the servers finishes its service, the age will drop again, resulting in successful absorption. Fig.~\ref{fig:amc_success_2} depicts a similar scenario, where we have considered that $S$ goes on vacation at time $t_1$. Therefore, in this scenario,  the ACM transitions to state $1$ at time $t_1$, and later when $S_n$ finishes service at time $t_n$, it will transition to state $6$. Both these scenarios illustrate successful absorptions and notice in both scenarios that once $P$ has been successful, the age curve coincides with the red dashed curve, which models the time spent till successful absorption. Therefore, in these two AMCs, the time spent during states 6 to 10 corresponds to the time duration of one AoI cycle, whereas the total time elapsed from the moment the AMC was initiated, is exactly equal to the age of the curve when the AMC resides in the same set of states.

In Fig.~\ref{fig:amc_unsuccessful}, we illustrate a scenario which results in an unsuccessful absorption. In here, at time $t_1$, we assume a new packet from source-$n$ enters $S$ and therefore the AMC transitions to state 3 at time $t_1$. Now, note that the packet in $S$ is having a later time stamp than the packet in $S_n$. Then, at time $t_2$,  when server $S$ finishes its service, the packet in $S_n$ will become obsolete. Therefore, when $S_n$ finally finishes its service at time $t_n$, the age will not drop and its packet is simply discarded by the remote monitor. Therefore, we consider that this AMC resulted in an unsuccessful absorption at time $t_2$. Note that, in this particular scenario, at time $t_n$, the age curve no longer aligns with the red dashed curve and hence this AMC does not model the age during an AoI cycle.
  
\begin{figure}[t]
    \centering
    \begin{subfigure}[b]{\columnwidth}
         \centering
         \includegraphics[scale=0.6]{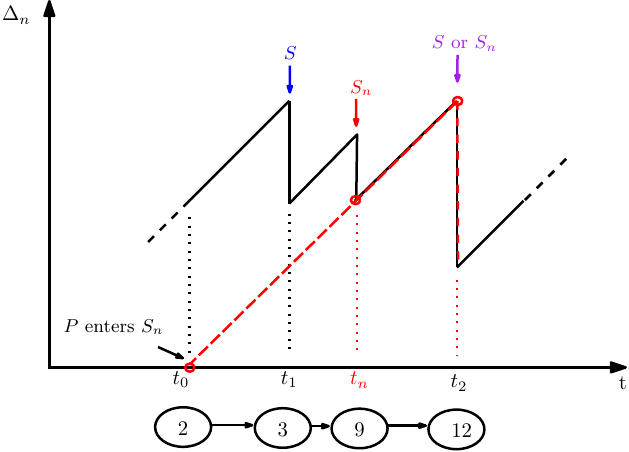}
         \caption{Successful absorption. }
         \label{fig:amc_success_1}
    \end{subfigure}
    \begin{subfigure}[b]{\columnwidth}
         \centering
         \includegraphics[scale=0.6]{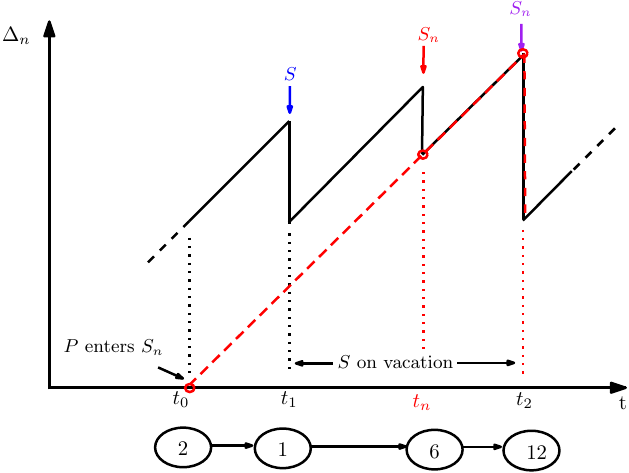}
         \caption{Successful absorption with intermediate vacation.}
         \label{fig:amc_success_2}
         \vspace{3mm}
    \end{subfigure}
     \begin{subfigure}[b]{\columnwidth}
         \centering
         \includegraphics[scale=0.6]{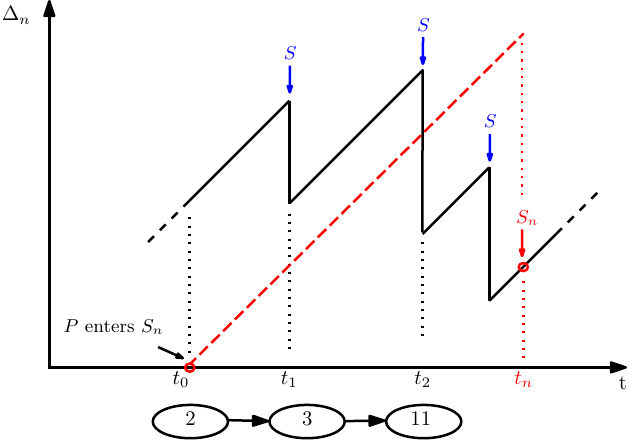}
         \caption{Unsuccessful absorption.}
         \label{fig:amc_unsuccessful}
    \end{subfigure}
    \caption{Successful absorption vs unsuccessful absorption.}
    \label{fig:AMC_examples}
\end{figure}

Next, we need to find the initial probability vector of the transient states of the AMC. For this purpose, we construct an RMC  whose states represent the states of the system from the perspective of a newly joining packet. Based on the timestamps of the packets in each server and whether the server $S$ is on vacation or not, we can define three states for  PS. These states are given in Table~\ref{tab:PS_RMC_states} and the transition rates of this RMC are presented in Fig.~\ref{fig:PS_RMC_states}.

\begin{table}[]
    \centering
    \caption{States of the RMC for the PS sub-problem.}
    \label{tab:PS_RMC_states}
    \begin{tabular}{|c|c|}
    \hline
         State& Description  \\
         \hline
         $R_1$& $S_n$ busy, $S$ on vacation \\
         \hline
         $R_2$& $S_n$ busy, $S$ busy, $T_n\geq T_s$ \\
         \hline
         $R_3$& $S_n$ busy, $S$ busy, $T_n < T_s$ \\
         \hline
    \end{tabular}
    
\end{table}

Let $\bm{\pi}=[\pi_1,\pi_2,\pi_3]$ be the stationary distribution of the above RMC. By algebraic manipulations, one can write $\bm{\pi}$ as,
\begin{align}
    \bm{\pi} = 
    \begin{bmatrix}
        q_n&\frac{p_n \mu_n}{\mu_n+\mu} & \frac{p_n\mu}{\mu_n+\mu}
    \end{bmatrix}.
\end{align}
Using $\bm{\pi}$, we can find $\bm{\sigma}$ as follows: Let the net rate at which a new source-$n$ packet joins the system be denoted by $f$. Note that a new packet enters the system with rate $\mu_n+p_n\mu$ when in any of the states of the RMC. Therefore, $f = \mu_n+p_n\mu$. Now, we establish the relation between the AMC and the RMC.
\begin{itemize}
    \item When in state $R_1$, a new source-$n$ packet will join server $S_n$ with rate $\mu_n$ and server $S$ with rate $p_n\mu$. The former event corresponds to the packet $P$ initiating the AMC from state 1 and the latter corresponds to initiating the AMC starting from state 4.
    \item When in state $R_2$ or $R_3$, similar to state $R_1$,  a new source-$n$ packet will join server $S_n$ with rate $\mu_n$ and server $S$ with rate $p_n\mu$. If the new packet joins $S_n$, it would correspond to the event that the packet $P$ initiated the AMC evolution starting from state 2 and if the new packet joins $S$, $P$ would start its AMC from state 4.
\end{itemize}

Based on the above observations, it is clear that the AMC would be kicked off only from states 1, 2 or 4. Then, using $\bm{\pi}$, $f$ and the established relations, the non-zero elements of the initial probability vector $\bm{\sigma}$ can be written as follows,
\begin{align}
    \sigma_1 &= \frac{\mu_n\pi_1}{f}, \\
    \sigma_2 & = \frac{\mu_n (\pi_1+\pi_2)}{f},\\
    \sigma_4 & = \frac{p_n\mu(\pi_1+\pi_2+\pi_3)}{f}.
\end{align}
Thus, we write the row vector $\bm{\sigma}$ explicitly as follows,
\begin{align}
    \bm{\sigma} =
    \begin{bmatrix}
        \frac{q_n\mu_n}{\mu_n+p_n\mu}&\frac{p_n\mu_n}{\mu_n+p_n\mu}&0&\frac{p_n\mu}{\mu_n+p_n\mu}&0&0&0&0&0&0
    \end{bmatrix}.
\end{align}
Since states 6 through 10 are  preceded by the event that $P$, which initiated the AMC, was successfully received by the remote monitor, the transient vector $\bm{\theta}$ can be written as follows,
\begin{align}
    \bm\theta^{T} =
    \begin{bmatrix}
       0&0&0&0&0&1&1&1&1&1
    \end{bmatrix}.
\end{align}
Note that the states corresponding to the non-zero values of the vector $\bm\theta$ are essentially the states that exactly coincide with the AoI curve as shown in Fig.~\ref{fig:aoi_cycle}.

\begin{figure}[t]
    \centering
    \includegraphics[scale=0.7]{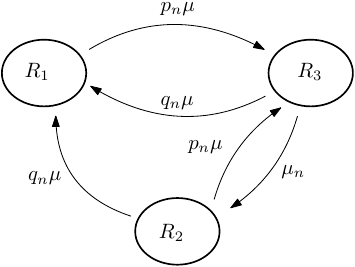}
    \caption{RMC for the PS sub-problem.}
    \label{fig:PS_RMC_states}
\end{figure}

Now, from Table \ref{tab:PS_rates}, we can obtain the sub-generator matrices $\bm U$ and $\bm{V}$ of the generator matrix of the AMC as follows,
\begin{align}
&\!\!\!\bm{U}  \!=\!
    \begin{bmatrix}
          * & 0 & p_n\mu&0&0&\mu_n&0&0&0&0\\
          q_n\mu & * & p_n\mu&0&0&0&\mu_n&0&0&0\\
          0&0&*&0&0&0&0&0&\mu_n&0\\
          0&0&0&*&\mu_n&0&0&p_n\mu&0&q_n\mu \\
          0&0&0&0&*&q_n\mu&0&0&p_n\mu&0\\
          0&0&0&0&0&*&0&0&p_n\mu&0\\
          0&0&0&0&0&q_n\mu&*&0&p_n\mu&0\\
          0&0&0&0&0&0&0&*&\mu_n&0 \\ 
          0&0&0&0&0&0&0&0&*&0\\
          0&0&0&0&0&\mu_n&0&p_n\mu&0&*
     \end{bmatrix}
    \\
    &\!\!\!\bm{V}^{T}  \!=\! 
    \begin{bmatrix}
          0&0&\mu&0&\mu_n&0&0&0&0&0\\
          0&0&0&0&0&\mu_n&\mu_n&\mu&\mu_n+\mu&0
    \end{bmatrix}, 
\end{align}
where the negative diagonal elements (represented by $*$) of $\bm{U}$ are chosen so as to satisfy $\bm{U} \bm{1}+\bm{V} \bm{1}= \bm{0}$ and $\bm{1}$ is a vector of ones of appropriate size. Having obtained the matrices $\bm{U}$ and $\bm{\sigma}$, we can now obtain the average age of source-$n$ from~\eqref{eqn:aoi_form}. Note that $\bm{U}$ is a block upper triangular matrix with two $5 \times 5$ blocks on the main diagonal. Therefore, the matrices $\bm{U}^{-1}$ and $\bm{U}^{-2}$ in~\eqref{eqn:aoi_form} can be obtained by inverting the $5 \times 5$ block matrices at the main diagonal only. Moreover, each of these two blocks has two zero columns except for a non-zero diagonal entry. Using this special structure of these blocks along with simple algebraic manipulations, the average age of source-$n$ can be written in the following closed form, 
\begin{align}
    \e[\Delta_n]= \frac{p_n^2\mu^2(2\mu_n+\mu)+p_n\mu(2\mu_n+\mu)^2+2\mu_n(\mu_n+\mu)^2}{(\mu_n+\mu)^2(p_n\mu+\mu_n)^2}.\label{eqn:PS_age_n}
\end{align}

\subsection{Cyclic Scheduler AoI}
Now, we will present the AoI analysis for the CS sub-problem. Here, the shared server scheduling decisions will be based on a CS where at source-$n$ scheduling instances, the shared server will pull a packet from source-$n$ once it is free, and would be on vacation at other scheduling instances. Let $P$, $P_n$, $T_n$ and $T$ be as defined in Section~\ref{sec:ps_aoi}. Let $\tilde C=[c_1,c_2,\dots,c_m]$ be the binary cyclic pattern with $c_i \in  \{1,2\}$, where $c_i=1$ corresponds to scheduling a source-$n$ transmission and $c_i=2$ corresponds to taking a vacation.  

Next, we will describe the construction of the AMC for CS. For brevity, we will reuse the state-space described in Table~\ref{tab:PS_states} to define a two-dimensional state vector for the transient states of the  CS AMC. We define the transient states of this AMC as the pairs $(i,j)$, where $i\in \{1,2,\dots,m\}$ denotes the $i$th scheduling instance of the pattern $\tilde C$ and $j\in\{1,2\dots,10\}$ corresponds to one of the states that was described in  Table~\ref{tab:PS_states}.

Note that  $c_i=1$ implies that the shared server $S$ is currently occupied by a source-$n$ packet, and therefore, $j$ can only take values in $\{2,3,4,5,7,8,9\}$. Similarly, $c_i=2$ implies that the shared server $S$ is currently on vacation, and therefore, $j$ can only take values in $\{1,6,10\}$. As before, let the unsuccessful and successful absorbing states be denoted by states 11 and state 12, respectively. Thus, altogether we will have $7k+3(m-k)+2$ states for the AMC, where $k$ is the number of source-$n$ occurrences within the pattern $\tilde C$. The transition rates of the CS AMC is summarized in Table~\ref{tab:CS_rates}, where $c_{m+1}$ is treated as $c_1$. The details of the state transitions are similar to the PS sub-problem with the only exception being that whenever the shared server finishes its transmission, $i$ changes from $i \to i+1$ ($m$ changes to $1$).

\begin{table}[]
    \centering
    \caption{Transition rates for the CS sub-problem.}
    \label{tab:CS_rates}
    \begin{tabular}{|c|c|c|c|}
    \hline
    \multicolumn{4}{|c|}{Transition rates}\\
    \hline
    $c_i$&From&To& Rate \\
    \hline
    \multirow{17}{*}{1}&\multirow{3}{*}{$(i,2)$}& $(i,7)$ & $\mu_n$\\
        \cline{3-4}
        &&$(i+1,3)$ if $c_{i+1}=1$& $\mu$\\
        \cline{3-4}
        &&$(i+1,1)$ if $c_{i+1}=2$& $\mu$\\
        \cline{2-4}
    &\multirow{2}{*}{$(i,3)$}& $(i,9)$ & $\mu_n$\\
        \cline{3-4}
        &&11& $\mu$\\
        \cline{2-4}
    &\multirow{3}{*}{$(i,4)$}& $(i,5)$ & $\mu_n$\\
        \cline{3-4}
        &&$(i+1,8)$ if $c_{i+1}=1$& $\mu$\\
        \cline{3-4}
        &&$(i+1,10)$ if $c_{i+1}=2$& $\mu$\\
        \cline{2-4}
    &\multirow{3}{*}{$(i,5)$}& 11  & $\mu_n$\\
        \cline{3-4}
        &&$(i+1,9)$ if $c_{i+1}=1$& $\mu$\\
        \cline{3-4}
        &&$(i+1,6)$ if $c_{i+1}=2$& $\mu$\\
        \cline{2-4}
     &\multirow{3}{*}{$(i,7)$}& 12  & $\mu_n$\\
        \cline{3-4}
        &&$(i+1,9)$ if $c_{i+1}=1$& $\mu$\\
        \cline{3-4}
        &&$(i+1,6)$ if $c_{i+1}=2$& $\mu$\\
        \cline{2-4}      
     &\multirow{2}{*}{$(i,8)$}& $(i,9)$ & $\mu_n$\\
        \cline{3-4}
        &&12& $\mu$\\
        \cline{2-4}
     &\multirow{1}{*}{$(i,9)$}& 12 & $\mu_n+\mu$\\
        \cline{1-4}
    \multirow{9}{*}{2}&\multirow{3}{*}{$(i,1)$}& $(i,6)$ & $\mu_n$\\
        \cline{3-4}
        &&$(i+1,3)$ if $c_{i+1}=1$& $\mu$\\
        \cline{3-4}
        &&$(i+1,1)$ if $c_{i+1}=2$& $\mu$\\
        \cline{2-4}
    &\multirow{3}{*}{$(i,6)$}& 12  & $\mu_n$\\
        \cline{3-4}
        &&$(i+1,9)$ if $c_{i+1}=1$& $\mu$\\
        \cline{3-4}
        &&$(i+1,6)$ if $c_{i+1}=2$& $\mu$\\
        \cline{2-4}
    &\multirow{3}{*}{$(i,10)$}&$(i,6)$   & $\mu_n$\\
        \cline{3-4}
        &&$(i+1,8)$ if $c_{i+1}=1$& $\mu$\\
        \cline{3-4}
        &&$(i+1,10)$ if $c_{i+1}=2$& $\mu$\\
        \cline{1-4}                    
    \end{tabular}
    
\end{table}

Next, we give the states of the RMC to compute the initial probabilities for the above AMC. Again, we will reuse the states of the RMC from the PS to define a two dimensional state vector for RMC of the CS. Let $(i,R_j)$ be the states of the RMC, where $i\in\{1,2,\dots,m\}$ represents the $i$th scheduling instance of $\tilde C$ and $R_j$ for $j =1$ to 3 denote the status of the packets as defined in Table~\ref{tab:PS_RMC_states}. Here, $c_i=1$ indicates that the shared server $S$ is occupied by a source-$n$ pattern, and therefore, the packet states can either be in $R_2$ or $R_3$. If $c_i=2$, then that indicates that the $S$ is currently on vacation, and therefore, packet status can only correspond to $R_1$. Thus, the CS RMC comprises $2k+(m-k)$ recurrent states in total. The associated transition rates of the RMC are given in Table~\ref{tab:CS_RMC_rates}.

\begin{table}[]
    \centering
     \caption{Transition rates of the CS RMC.}
    \label{tab:CS_RMC_rates}
    \begin{tabular}{|c|c|c|c|}
        \hline
        \multicolumn{4}{|c|}{Transition states}\\
        \hline
        $c_i$ &To & From & Rate\\
        \hline
        \multirow{5}{*}{1}&\multirow{2}{*}{$(i,R_2)$}& $(i+1,R_3)$ if $c_{i+1}=1$&$\mu$\\
            \cline{3-4}
        &&$(i+1,R_1)$ if $c_{i+1}=2$& $\mu$\\
        \cline{2-4}
        &\multirow{2}{*}{$(i,R_3)$}& $(i,R_2)$ &$\mu_n$\\
        \cline{3-4}
        &&$(i+1,R_3)$ if $c_{i+1}=1$& $\mu$\\
        \cline{3-4}
         &&$(i+1,R_1)$ if $c_{i+1}=2$& $\mu$\\
        \cline{1-4}
        \multirow{2}{*}{2}&\multirow{2}{*}{$(i,R_1)$}& $(i+1,R_3)$ if $c_{i+1}=1$&$\mu$\\ 
         \cline{3-4}
         &&$(i+1,R_1)$ if $c_{i+1}=2$& $\mu$\\
        \cline{1-4}
    \end{tabular}
   
\end{table}

Now, we need to find the total rate $f_c$ at which a new packet enters the system. Note that, if we are in state $(i,R_j)$ and $c_{i+1}=1$, then a new source-$n$ packet would enter the system with rate $\mu+\mu_n$, and if $c_{i+1}=2$, then the rate would be $\mu_n$. Let $\phi_{i,j}$ denote the stationary probability of the state $(i,R_j)$. Then, we can write the quantity $f_c$ as follows,
\begin{align}
    f_c = \sum_{(i,j)}\phi_{i,j}(\mu_n+\mu\mathbf{1}_{\{c_{i+1}=1\}}),
\end{align}
where the summation is over the state-space of the RMC, $c_{m+1}=c_1$ and $\mathbf{1}_{\{\cdot\}}$ is the indicator function. The corresponding relations between the AMC and RMC for the CS are as follows:
\begin{itemize}
    \item  When in state $(i,R_1)$ and if $c_{i+1}=1$, a newly joining source-$n$ packet entering server $S_n$ would correspond to initiating the AMC starting from state $(i,1$). If the packet joins the shared server $S$ instead, then the AMC will be initiated starting from state $(i+1,4)$. These events would occur with rates $\mu_n$ and $\mu$, respectively. If $c_{i+1}=2$, then only the former event would occur.
    \item When in state $(i,R_2)$ or $(i,R_3)$, if the new source-$n$ packet joins server $S_n$, then the AMC would start from state $(i,2)$ and this event occurs with rate $\mu_n$. If $c_{i+1}=1$, and the packet joins server $S$, then the AMC would start from state $(i+1,4)$ and this event occurs with rate $\mu$.
\end{itemize}

The above relations indicate that the AMC can only start evolving from the states $(i,2)$, $(i,4)$ and $(i,1)$. Let $\bm{\sigma}=\{\sigma_{i,j}\}$ represent the initial probability vector of the CS AMC, where $\sigma_{i,j}$ denotes the probability of being in state $(i,j)$ of the CS AMC. Using the above relations and the rate $f_c$, we can define the initial probabilities of the CS AMC as follows,
\begin{align}
    \sigma_{i,j}=
    \begin{cases}
        \frac{\mu_n\phi_{i,1}}{f_c}, &\text{if}\; c_i=2,j=1,\\
        \frac{\mu_n(\phi_{i,2}+\phi_{i,3})}{f_c}, & \text{if}\; c_i=1,j=2,\\
        \frac{\mu\phi_{i-1,1}}{f_c}, & \text{if}\; c_i=1, c_{i-1}=2,j=4,\\
        \frac{\mu(\phi_{i-1,2}+\phi_{i-1,3})}{f_c}, & \text{if}\; c_i=1,c_{i-1}=1,j=4,\\
        0, & \text{otherwise},
    \end{cases}
\end{align}
where $c_{0} = c_m$. The associated transient vector $\bm{\theta}$ will be given by,
\begin{align}
    \theta_{i,j}=
    \begin{cases}
        1, & \text{if}\; c_i=1, j\in \{7,8,9\},\\
        1, & \text{if}\; c_i=2, j\in \{6,10\},\\
        0, & \text{otherwise}.
    \end{cases}
\end{align}
Then, as before, by using \eqref{eqn:aoi_form}, we can find the average AoI of source-$n$ for the CS. However, since the matrices and vectors involved depend on the pattern considered, we do not provide a closed-form expression for this scheduler.

\section{Optimal Probabilistic Scheduler} \label{sec:opt_PS}
In this section, we pose the construction of the optimal PS as a convex optimization problem and obtain the optimal values for $p_n$. Using \eqref{eqn:aoi_form}, we obtain the expected weighted AoI as,
\begin{align}
    \e[\Delta] =  & \sum_nw_n\cdot\frac{(2\mu_n+\mu)\mu y_n+\mu\mu_n(\mu_n+\mu)}{(\mu_n+\mu)^2y_n^2} \nonumber\\
    &+\sum_n w_n\cdot\frac{(2\mu_n+\mu)}{(\mu_n+\mu)^2},
\end{align}
where  $y_n=p_n\mu+\mu_n$. Since the first summation is a linear sum of functions $\frac{1}{y_n}$ and $\frac{1}{y_n^2}$ which are both convex for $y_n\ge0$, the weighted AoI is a convex function with respect to $y_n$'s. Therefore, finding the optimal probabilities reduces to the following convex problem:
\begin{mini}
    {y_n}{\sum_n \left(\frac{(2\mu_n+\mu)\mu}{(\mu_n+\mu)^2}\cdot \frac{w_n}{y_n}+\frac{\mu\mu_n}{(\mu_n+\mu)}\cdot\frac{w_n}{y_n^2} \right)}
    {\label{Optimization1}}
    {}
    \addConstraint{\sum_n y_n}{= \mu + \sum_n\mu_n}
    \addConstraint{y_n}{\geq \mu_n}.
\end{mini}

\begin{algorithm}[t]
    \caption{An algorithm to find the optimal PS}\label{alg:cap}
    \begin{algorithmic}[1]
    \Require $\mu$, $\{w_n$, $\mu_n\}_{n=1}^N$, $\lambda_u$ sufficiently large, $\lambda_l$, $\epsilon$ sufficiently small
    \State $a_n =  \frac{w_n(2\mu_n+\mu)\mu}{(\mu_n+\mu)^2}$, $b_n =\frac{\mu\mu_n}{(\mu_n+\mu)} \quad \forall n $
    \State $\eta = \mu+\sum_n \mu_n $
    \State $y_n=0 \quad \forall n$
    \While{$|\eta - \sum_ny_n|>\epsilon$}
    \State $\lambda= \frac{\lambda_u+\lambda_l}{2}$
    \State $f_n(y) \coloneq \lambda y^3-a_ny-2b_n \quad \forall n$
    \State $z_n \gets $ positive real root of  $f_n(y_n)=0$ 
    \State $y_n= \max\{z_n,\mu_n\}$
    \If{$\sum_n y_n > \eta$}
        \State $\lambda_l=\lambda$
    \Else
        \State $\lambda_u=\lambda$
    \EndIf
    \EndWhile
    \State{\textbf{Output:}} $p_n = \frac{y_n-\mu_n}{\mu} \quad \forall n$
    \end{algorithmic}
    \label{alg:opt_prob}
\end{algorithm}

Let $a_n$ and $b_n$ denote the coefficients of $\frac{1}{y_n}$ and $\frac{1}{y_n^2}$ terms in the objective function, respectively. We define the Lagrangian of the above optimization problem as follows,
\begin{align}
    \!\!\!\mathcal{L}=\sum_n\frac{a_n}{y_n}+\frac{b_n}{y_n^2} +\lambda\Big(\sum_ny_n-\eta\Big)+\sum_n\gamma_n(\mu_n-y_n), \!\!
\end{align}
where $\bm{y}=\{y_1,\dots,y_n\}$, $\eta =\mu +\sum_n \mu_n$, and $\{\gamma_1, \dots, \gamma_n\}$, $\lambda$ are the Lagrange multipliers of the inequality and equality constraints, respectively.  Since this satisfies the Slater's condition, the Karush-Kuhn-Tucker (KKT) conditions will yield the following sufficient conditions for optimality \cite{boyd},
\begin{align}
    a_ny_n+2b_n &= (\lambda-\gamma_n)y_n^3, \label{eqn:opt_1}\\
    \gamma_n(\mu_n-y_n) &= 0, \label{eqn:opt_2} \\
    \sum_n y_n &= \eta. \label{eqn:opt_3}
\end{align}
Since $\gamma_n\geq0$ and  any feasible $y_n$  satisfies $y_n\geq \mu_n$, we can conclude that $\lambda>0$. From \eqref{eqn:opt_2}, we have that if $\gamma_n>0$, then $y_n = \mu_n$ and if $y_n>\mu_n$, then $\gamma_n$ will be zero. If $\gamma_n=0$, then note that for a fixed $\lambda$, the $y_n$ that satisfies \eqref{eqn:opt_1} is simply the intersection of the straight line $a_ny_n+2 b_n$ and the cubic polynomial $\lambda y_n^3$. Since $\lambda, a_n, b_n>0$, a simple geometric interpretation reveals that \eqref{eqn:opt_1} has only one positive real root in this particular scenario. Moreover, this root will decrease as we increase $\lambda$ and can be found using Cardano's formula \cite{cardano}. To find the optimal $\lambda$, we can use a simple bisection search. Algorithm~\ref{alg:opt_prob} gives the pseudo-code for finding the optimal probabilities using the above steps.

Next, we present an important relationship between the optimum probabilistic and cyclic schedulers in Theorem~\ref{thrm:ps_v_cs} whose proof is given in Appendix~\ref{AppendixA}.
\begin{theorem}\label{thrm:ps_v_cs}
    The optimal cyclic scheduler achieves a lower weighted AoI than the optimal probabilistic scheduler.
\end{theorem}

\section{Cyclic Scheduler Construction}\label{sec:cyc}
In this section, we present two heuristic approaches for the construction of the cyclic scheduler where in one we refurbish the Insertion Search (IS) algorithm introduced in \cite{gau23} and in the other we utilize the optimal probabilities computed in Section~\ref{sec:opt_PS}.

\subsection{Insertion Search (IS) Algorithm}
Variants of the IS algorithm have been readily used in the past and they have been shown to perform well for cyclic scheduler construction to minimize the weighted AoI \cite{gau23, akar_etal_infocom24, NOTS2024}. In this paper, we use the IS algorithm with slight modifications to fit to our setting. The IS algorithm constructs the cyclic schedule in an iterative manner by improving upon the current best schedule obtained at each iteration. In the first iteration, we insert the source that would induce the lowest weighted AoI into our empty schedule. This is a key change from the IS algorithms employed in prior works for single-server scenarios where the IS algorithm is always initiated starting from a round robin (RR) schedule instead of an empty schedule. At each iteration, we select a new source to add to our current schedule that would result in the lowest weighted AoI. While doing so, we need to select a new position in the current schedule to add the new source instance. Hence, at each iteration, we select a source-position pair that achieves the lowest weighted AoI (among all the pairs) to construct a new cyclic schedule. Then, this new cyclic schedule is passed on to the next iteration where the same steps are repeated. At any iteration, if the weighted AoI does not improve, we will continue to explore $\ell$ more iterations (an exploration of $\ell=6$ iterations was used in the numerical results) and stop if no further improvement is observed. Then, we select the best schedule from among all the cyclic schedules that were obtained at each iteration.}

\subsection{Probability Aided Cyclic (PAC) Scheduler}
Despite the versatility of the IS algorithm, it comes with a high computational complexity.  Within each iteration of IS, the weighted AoI needs to be computed several times and this involves inverting a large matrix each time. This limits its applicability for systems with relatively large number of sources. To overcome this drawback, we introduce the so-called PAC scheduler which is a cyclic scheduler that is constructed using the optimal probabilities $p_n^*$ of PS. In the PAC scheduler, we aim to construct the cyclic schedule such that the proportion of time dedicated for source-$n$ tallies with $p_n^*$ of the optimal PS. Next, we select a schedule length $K$ and subsequently find the number of instances $K_n$ that should be allocated for source-$n$ such that $\sum_n K_n = K$. Then, we try to uniformly spread the source instances within the schedule to construct the PAC scheduler. Both these problems, i.e., finding $K$, $K_n$ and uniformly spreading the scheduling instances, have been studied in the past. The reference \cite{akar_etal_infocom24} introduces a deficit round robin (DRR) algorithm which constructs an almost uniform schedule given the optimal source frequencies within the schedule. Thus, by focusing only on the sources with $p_n^*>0$ and  replacing the optimal frequencies with $p_n^*$, we propose to use the DRR algorithm introduced in \cite{akar_etal_infocom24} to construct our PAC scheduler. For simulation purposes, we have used the same algorithm parameters of the SAMS-1 algorithm given in \cite{akar_etal_infocom24}.

\begin{figure}[t]
     \centering
     \includegraphics[width=0.8\columnwidth]{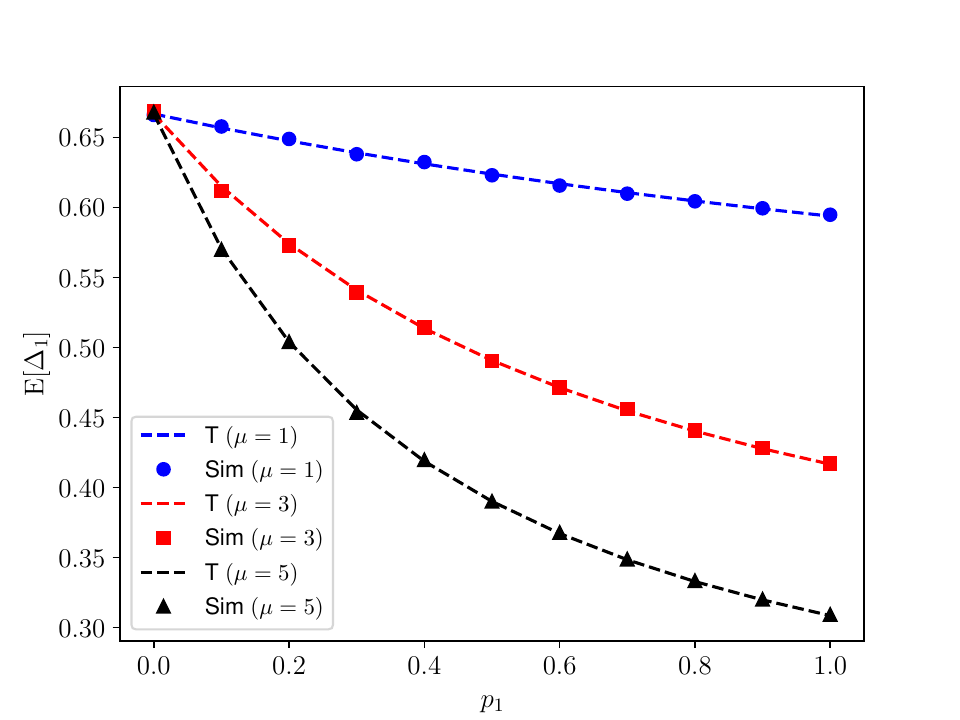}
     \caption{Theoretical values (T) vs simulated values (Sim) of $\e[\Delta_1]$ for PS with $\mu_1=3$. }
     \label{fig:PS_T_v_S}
 \end{figure}

  \begin{figure}[t]
     \centering
     \includegraphics[width=0.8\columnwidth]{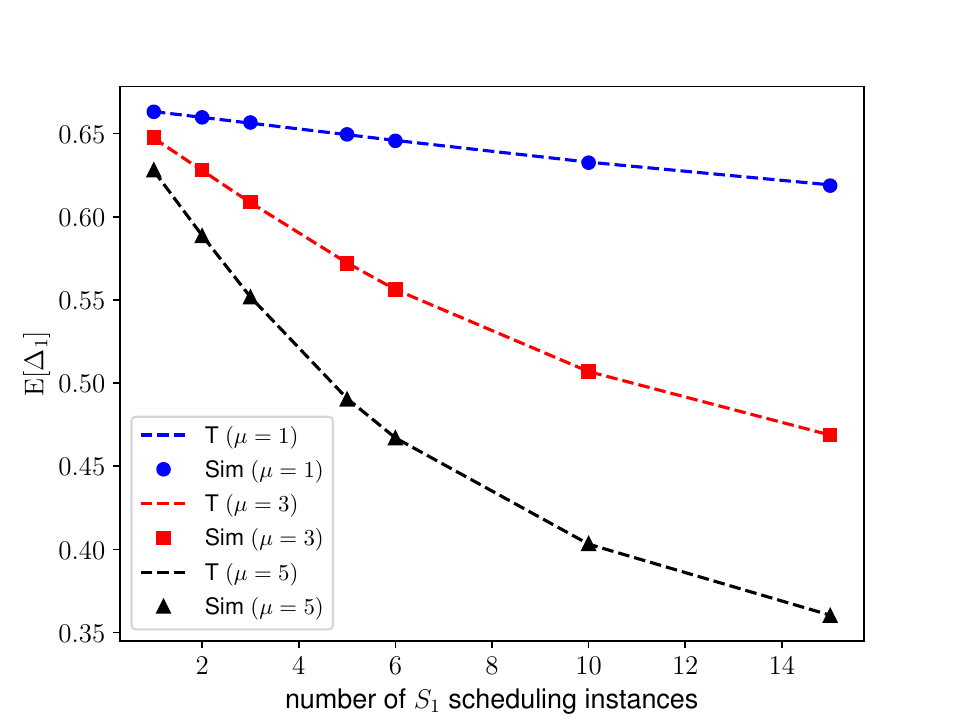}
     \caption{Theoretical values (T) vs simulated values (Sim) of $\e[\Delta_1]$ for CS with $\mu_1=3$ for cyclic patterns of length $30$ where $S_1$ scheduling instances are uniformly placed within the pattern.}
     \label{fig:CS_T_v_S}
 \end{figure}

\begin{figure}[t]
    \centering
    \includegraphics[width=0.8\columnwidth]{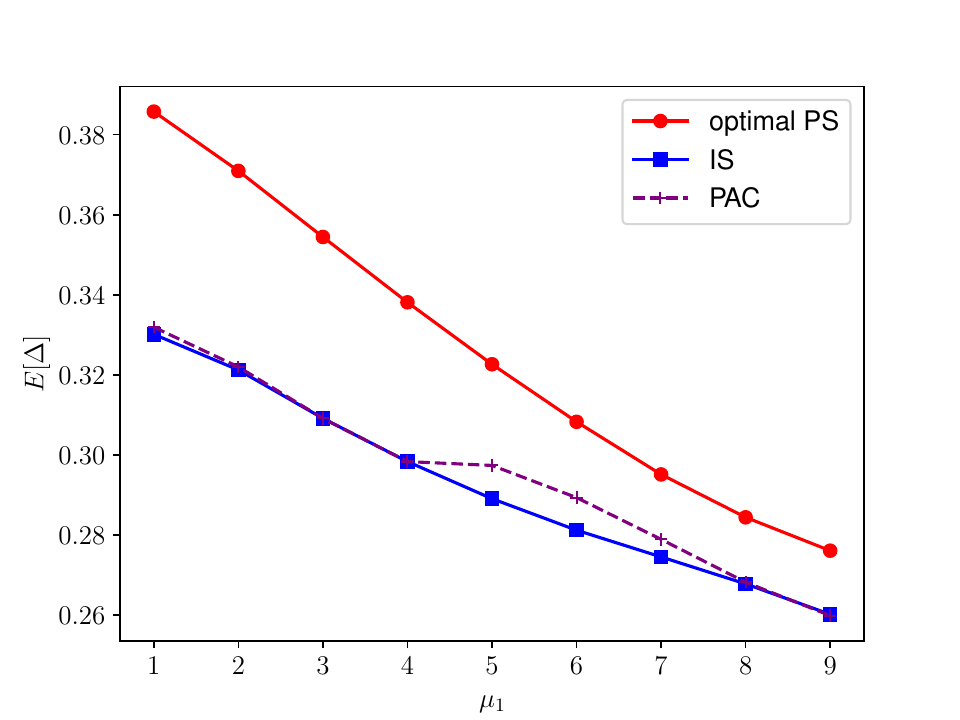}
    \caption{Variation of the weighted AoI with $\mu_1$ for $N=3$, $w_1 = 0.3$, $w_2=0.5$, $w_3=0.3$, $\mu=8$, $\mu_2=2$ and $\mu_3=3$.}
    \label{fig:mu_var}
\end{figure}

\begin{figure}[t]
    \centering
    \includegraphics[width=0.8\columnwidth]{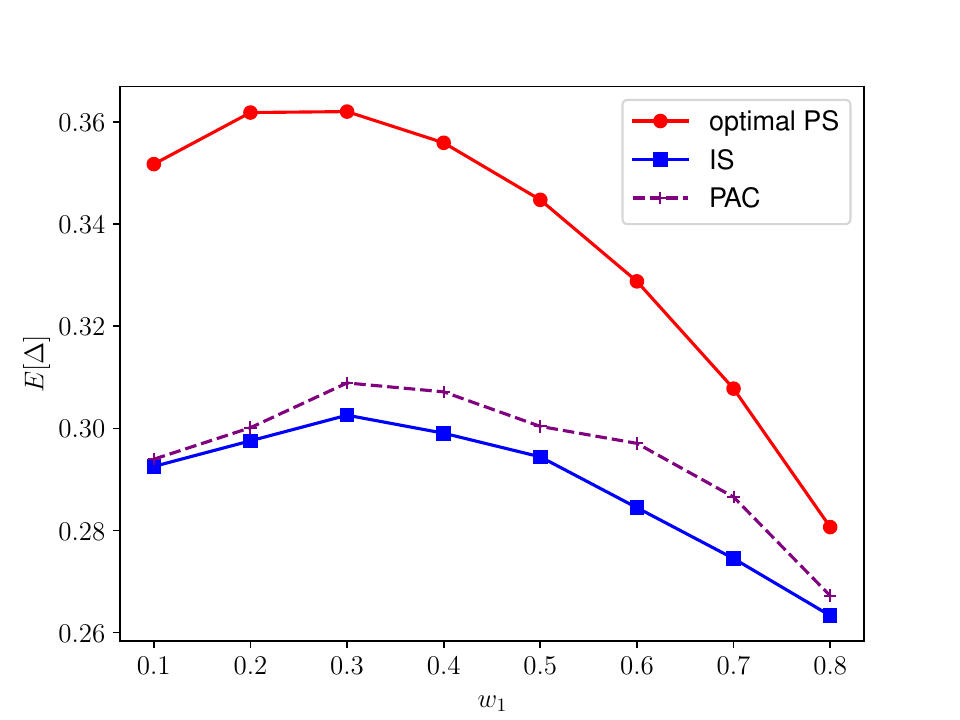}
    \caption{Variation of the weighted AoI with $w_1$ for $N=4$, $w_n = \frac{1-w_1}{3}, n \neq 1$, $\mu=10$ and $\mu_n=n$.}
    \label{fig:w_var}
\end{figure}

\section{Numerical Results}\label{sec:results}
In this section, we first validate our analytical expressions and then evaluate the performance of the three scheduling schemes, namely PS, IS, and PAC. We focus our attention to source-$1$ in Fig.~\ref{fig:PS_T_v_S} which depicts the theoretical and simulated values of $\e[\Delta_1]$ for PS with three choices of $\mu$ for PS, as we vary $p_1$. We observe that the simulation values closely follow the expression given \eqref{eqn:PS_age_n}. On the other hand, Fig.~\ref{fig:CS_T_v_S} depicts the comparison between the theoretical and simulated values for CS as we vary the number of scheduling instances of source-$1$ within a pattern of length $30$ that are spread out uniformly within the pattern. Again, three values of $\mu$ are considered. Fig.~\ref{fig:PS_T_v_S} and Fig.~\ref{fig:CS_T_v_S} demonstrate that the simulated values of $\e[\Delta_1]$ closely align with the theoretical values for both PS and CS obtained with the proposed analytical method, validating the correctness of our analysis.

In the second experiment, we consider a system with three sources for which we fix the weights given to the sources, and we vary the dedicated service rate of source-$1$ when $\mu=8$, $\mu_2=2$ and $\mu_3=3$. As shown in Fig.~\ref{fig:mu_var}, cyclic scheduling schemes outperform the optimal PS, and IS has a slight edge over the PAC scheduler. This validates the rationale behind selecting the optimal scheduling probabilities as a suitable substitute for the optimal source frequencies for the construction of the cyclic scheduler in PAC.

In the next experiment, we consider a system of four sources, where we fix the service rates of the sources by setting $\mu=10$ and $\mu_n=n,\   1 \leq n \leq 4$. We also vary the weight $w_1$ assigned to the first source when $w_n = \frac{1-w_1}{3}, n \neq 1$. The variation of the weighted AoI for this scenario is shown in Fig.~\ref{fig:w_var}. As depicted, cyclic scheduling schemes exhibit superior AoI performance compared to the probabilistic scheme. Moreover, we observe that, the weighted AoI first increases and then starts to decrease with $w_1$. This is because, as $w_1$ increases initially, its contribution to the weighted AoI increases, but as $w_1$ increases further, the shared channel tends to favor source-$1$ more, and hence, reduces the average AoI of source-$1$, which in turn reduces the weighted AoI of the system.

\begin{figure}[t]
    \centering
    \includegraphics[width=0.8\columnwidth]{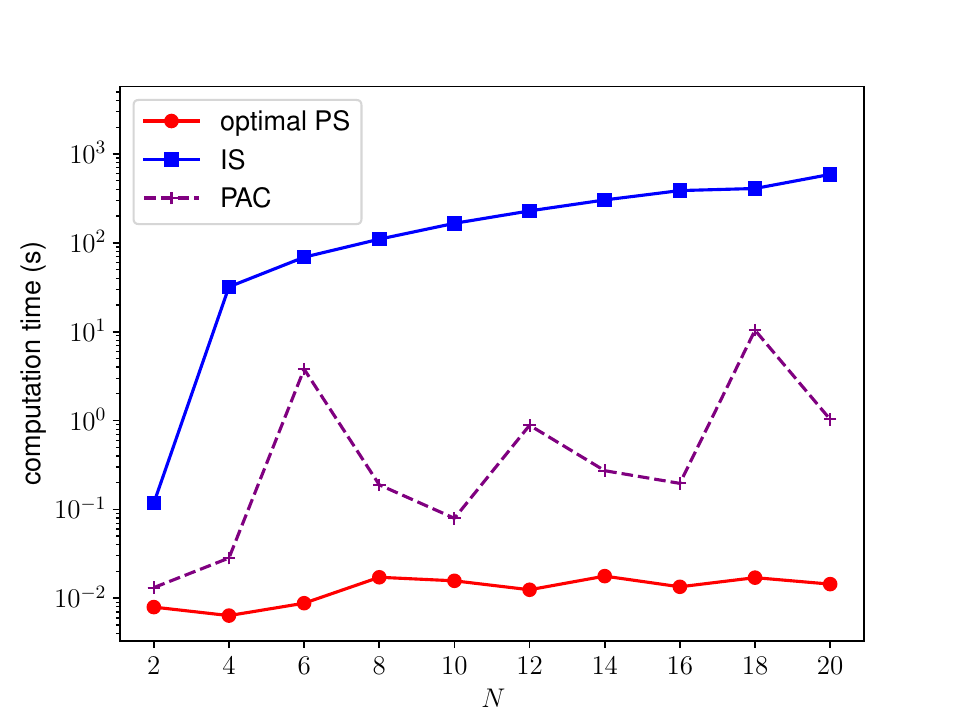}
    \caption{Variation of the computational time in seconds (log scale) with $N$ for $w_n = \frac{1}{N}$, $\mu=\frac{N}{2}$, $\mu_1=N$, $\mu_n=n$, $n\geq 2$.}
    \label{fig:comp_time}
\end{figure}

\begin{figure}[t]
    \centering
    \includegraphics[width=0.8\columnwidth]{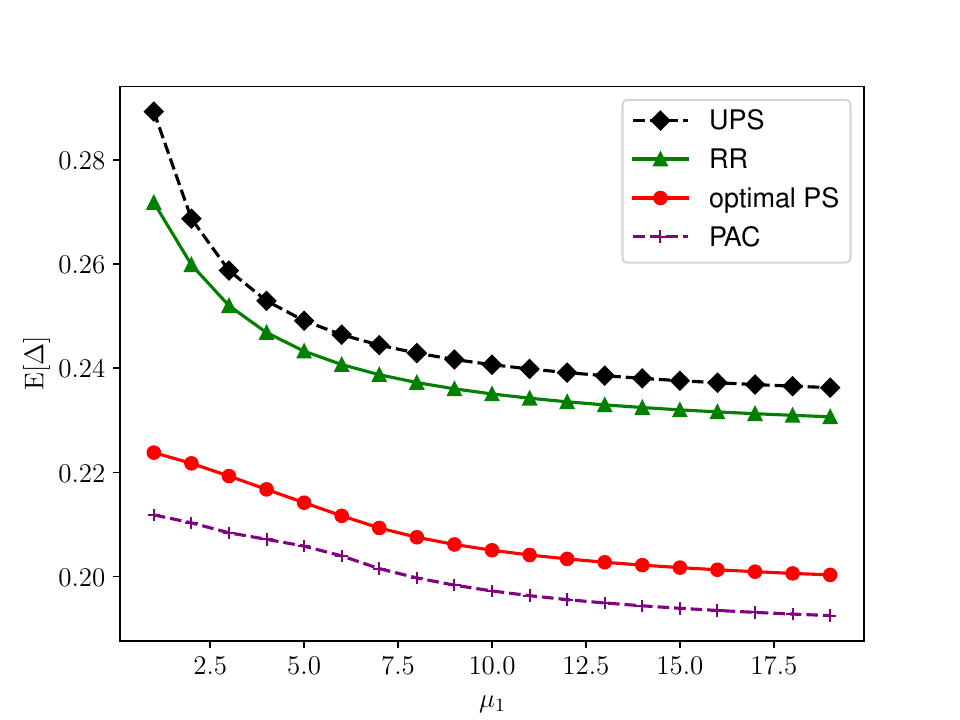}
    \caption{Variation of the weighted AoI with $\mu_1$ for $N=20$, $w_n = \frac{1}{20}$, $\mu=10$, $\mu_n=n$, $n\geq 2$.}
    \label{fig:LS_var}
\end{figure}

Despite the efficacy of the IS algorithm, it is limited by its computational complexity. However, the PAC algorithm is readily applicable to any number of sources. To demonstrate this, the variation of the execution times with respect to the number of sources for the three studied schemes is given in Fig.~\ref{fig:comp_time}. As depicted, the execution time of IS is considerably higher compared to PAC and PS. This makes the IS algorithm inconvenient for large-scale problems and for scenarios when the channel service rates (or the estimates of the service rates) change from time to time. For example, if we do not know the channel service times exactly, we may be able estimate them by observing the transmissions across a reasonable period of time. As our estimates improve, we may need to change the schedule. Since IS takes up a considerable portion of time to construct the schedule, our estimates of the channel rates may have changed. For such situations, having in hand a computationally efficient algorithm to construct the schedules could significantly improve the overall system performance. For a relatively large status update system with $N=20$ sources, we evaluate the performance of only the PS and the PAC schedulers where we vary the dedicated service rate of source-$1$ while assuming symmetric weights and $\mu=10$, $\mu_n=n$, $n\geq 2$. For comparison, we also consider two other naive scheduling policies termed uniform probabilistic scheduler (UPS) where $p_n= \frac{1}{N}, \ \forall n$ and the RR scheduler which follows a cyclic round robin pattern catering each source once within a cycle of $N$. The weighted AoI results are depicted in Fig.~\ref{fig:LS_var} demonstrating clearly the benefits of employing cyclic scheduling as opposed to probabilistic scheduling.

\begin{figure}[t]
    \centering
    \includegraphics[width=0.8\columnwidth]{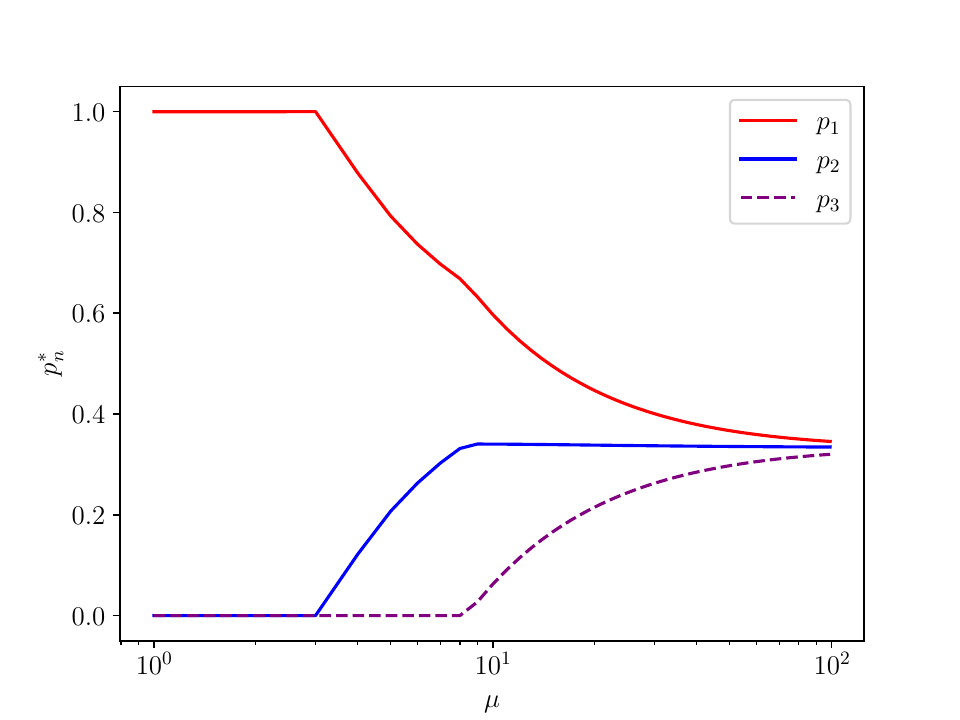}
    \caption{Variation of the optimal scheduling probabilities with $\mu$ (log scale) for $N=3$, $w_n=\frac{1}{3}$, $\mu_1=4$, $\mu_2= 7$ and $\mu_3=10$.}
    \label{fig:p_var}
\end{figure}

Next, we evaluate how the optimal probabilities of PS vary with the service rate of the shared server for a simple three-source system whose weights are the same but with different dedicated service rates. As shown in Fig.~\ref{fig:p_var}, when the shared server rate is small compared to the dedicated server rates, the shared server tries to give more scheduling priority to the source with the lowest dedicated rate. However, as $\mu$ is comparatively large compared to the dedicated service rates, equal priority is given to all the sources. In this case, it is as if the dedicated servers are non-existent from the perspective of the shared server, and almost all packets served through the dedicated links will turn out to be obsolete. Thus, when $\mu$ is large when compared to the dedicated rates, our problem reduces to a single-server multi-source scheduling problem.

\begin{figure}[t]
    \centering
    \includegraphics[width=0.8\columnwidth]{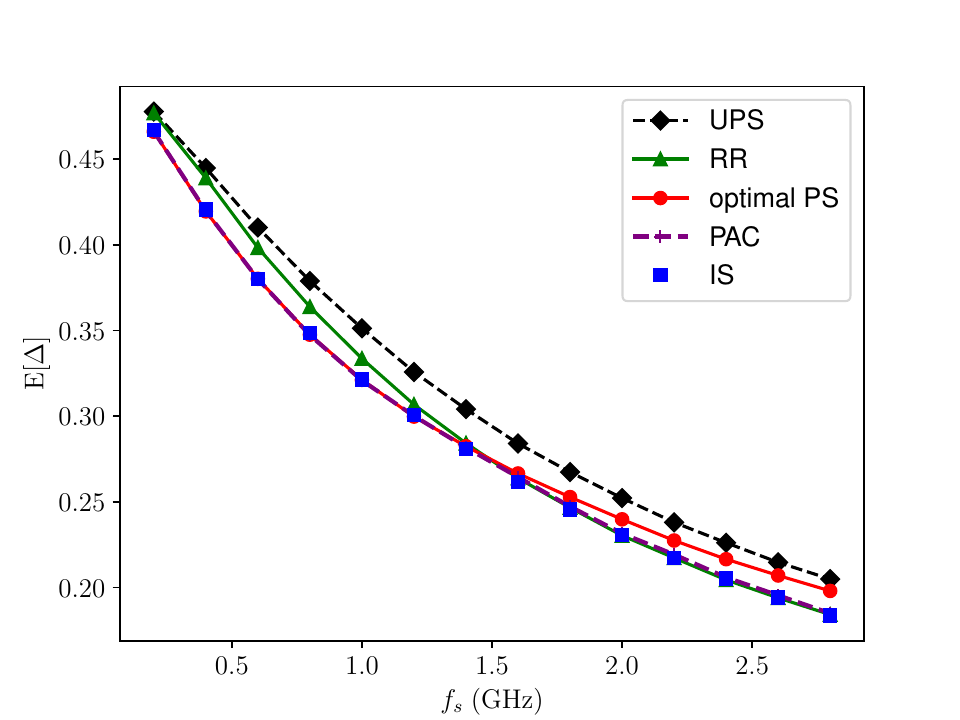}
    \caption{Variation of simulated values of the weighted AoI  with cloud server clock frequency $f_s$ for an IoT system with two edge devices.}
    \label{fig:iot_sim}
\end{figure}

Finally, we simulate and evaluate our policies for a mobile edge computing IoT application inspired by \cite{industrial_iot} where the IoT end devices offload their tasks to an edge server. Let $\text{TS}$ denote the average size of tasks processed by these devices. Let the clock frequency $f$, number of clock cycles per instruction (CPI) and the number of bits per instruction (BPI) for the $i$th end device be denoted by $f_i$, $\text{CPI}_i$ and $\text{BPI}_i$, respectively. Similarly, let $f_s$, $\text{CPI}_s$ and $\text{BPI}_s$, denote the same set of parameters corresponding to the edge server. Now, let us consider the following scenario where we assume $\text{TS}=50\text{MB}$ ($1\text{M}$ taken as $2^{20}$) and the edge server employs a CPU with 32 bit instruction set architecture (ISA) with $\text{BPI}_s=32$ and $\text{CPI}_s= 20$. This server is assumed to be shared among the two IoT devices 1 and 2, each of which employs a CPU with a 16 bit ISA with 5 and 7 CPI, respectively, with $\text{BPI}_i=16, \ i=1,2$, $\text{CPI}_1=5$, $\text{CPI}_2=7$, $f_1=1\text{GHz}$ and $f_2=0.5\text{GHz}$. Further, we assume that each device is given equal weights while processing their tasks with exponentially distributed service times with mean given by $(\text{TS}\times \text{CPI})/(\text{BPI}\times f)$. Fig.~\ref{fig:iot_sim} depicts the weighted AoI of various scheduling policies as we vary the clock frequency $f_s$ of the edge server. As illustrated, IS and PAC outperform all the other three policies at all edge server clock frequencies, whereas for this particular example, these two cyclic schedulers perform alike. Also note that, in this example, we focus on scheduling of computation servers and not of communication links, which can also be tackled with our proposed framework.

\section{Conclusion}\label{sec:con}
We provided an absorbing Markov chain (AMC) formulation to derive expressions for the weighted AoI of a status update system comprising multiple dedicated servers along with a single shared server. Through rigorous convex optimization, we have provided the optimal probabilistic scheduler along with several heuristics for developing cyclic schedulers which are shown to be superior in performance in comparison to several baseline schemes. As discussed in the numerical results section, when the service rate of the shared server is high, the packets through the dedicated servers would turn out to be obsolete most of the time. Therefore, the problem of selecting an appropriate service rate for the shared server so as to improve efficiency of the system (i.e., to minimize the number of obsolete packets) is an interesting future research direction for the system studied in this work. Moreover, the analysis of the introduced system model under random arrivals and also heterogeneous service times for the shared server, is another potential future work item. Analysis of the system in the presence of multiple shared servers and the development of scheduling algorithms that take into consideration additional performance criteria including fairness, power consumption, etc., are also left for future work.

\appendices
\section{Proof of Theorem \ref{thrm:ps_v_cs}}
\label{AppendixA}
 To prove the result, we will use a sample path argument on the scheduling instances of the PS. Let the optimal probabilities of the PS be denoted by $p_n^*$. At each scheduling instance, the scheduler will choose one of the sources according to the probabilities $p_n^*$ and this sequence of source allocations can be treated as the sample path of the scheduler. At any given scheduling instance, for each source, if more than two transmissions occur through the dedicated server before the transmission through the shared server is over, then regardless of which source the shared server was catering, the distribution of the age for that particular source at the beginning of the next scheduling instance will be the same. Let us call this event as the \emph{reset} event. Now, let us analyze the age of source-$n$ considering this reset event.
 
 For each source, at each scheduling instance, this reset event occurs with a fixed probability. Let $t$ be the number of scheduling instances till a reset event occurs for a given source. Then, $\exists  \ T\in \mathbb{N}$ such that $\mathbb{P}(t>T)<\epsilon$ for all sources. Moreover, $\exists \ L\in \mathbb{N}$ such that $\frac{T}{L}<\epsilon$. Now, let us look at the sample path of the scheduler and partition it into sequences of length $L$. Let these sequences be indexed as $C^{(l)}$. Now, we will compute the average age of the optimal PS using the traditional graphical method. The average age of the $n$th source will be given by,
 \begin{align}
     \e[\Delta_n]=\frac{\sum_lA_l^{(n)}\beta_l}{\sum_l D_l\beta_l},\label{eqn:ps_seq}
 \end{align}
 where $A_l^{(n)}$ is the expected area under the age curve of source-$n$ for sequence $C^{(l)}$, $D_l$ is the expected time duration for the sequence $C^{(l)}$ and $\beta_l$ is the probability of observing the sequence $C^{(l)}$ in the sample path of the scheduler. Note that since all the sequences are of length $L$, then $\sum_l D_l\beta_l=L\e[Y]$ where $\e[Y]=\frac{1}{\mu}$. Then, \eqref{eqn:ps_seq} can be simplified as follows,
 \begin{align}
      \e[\Delta_n]=\sum_l\beta_l\frac{A_l^{(n)}}{L\e[Y]}.
 \end{align}
 Let $\hat{A}_l^{(n)}$ be the expected area under the age curve of source-$n$ during the first $T$ slots of the sequence $C_l^{(n)}$, and $\tilde{A}_l^{(n)}$ be the expected area under the age curve of source-$n$ during the next $L-T$ slots of the sequence $C_l^{(n)}$. Then, $A_l^{(n)}=\hat{A}_l^{(n)}+\tilde{A}_l^{(n)}$. This gives us,
 \begin{align}
     \frac{A_l^{(n)}}{L\e[Y]}=\left(\frac{T}{L}\right)\frac{\hat{A}_l^{(n)}}{T\e[Y]}+\left(1-\frac{T}{L}\right)\frac{\tilde{A}_l^{(n)}}{(L-T)\e[Y]}.
 \end{align}
 Now, let us define $B_l^{(n)}$, $\hat{B}_l^{(n)}$ and $\tilde{B}_l^{(n)}$ as the counterparts of $A_l^{(n)}$, $\hat{A}_l^{(n)}$ and $\tilde{A}_l^{(n)}$, respectively, when instead of PS, CS with schedule $C^{(l)}$ is used. Then, the following holds,
 \begin{align}
     \left|\frac{A_l^{(n)}}{L\e[Y]}-\frac{B_l^{(n)}}{L{\e[Y]}}\right|\leq &\left(\frac{T}{L}\right)\left[\frac{\hat{A}_l^{(n)}}{T\e[Y]}+\frac{\tilde{A}_l^{(n)}}{(L-T)\e[Y]}\right]\nonumber\\
     &+\left(\frac{T}{L}\right)\left[\frac{\hat{B}_l^{(n)}}{T\e[Y]}+\frac{\tilde{B}_l^{(n)}}{(L-T)\e[Y]}\right]\nonumber\\
     &+\left|\frac{\tilde{A}_l^{(n)}}{(L-T)\e[Y]}-\frac{\tilde{B}_l^{(n)}}{(L-T)\e[Y]} \right|.
 \end{align}
Let $\tilde{\Delta}_n$ denote the average AoI of source-$n$, in the absence of the shared server. Since the age decreases in the presence of the shared server, we have that the first two terms of the above inequality to be bounded by $2\tilde{\Delta}_n$. Now, to bound the third term, we take into account the reset event for source-$n$. Let $t$ be the number of scheduling instances starting from the beginning of the sequence $C^{(l)}$ till a reset event occurs for source-$n$. Now, based on the reset event, we can further break down $\tilde{A}_l^{(n)}$ and $\tilde{B}_l^{(n)}$ terms as follows,
\begin{align}
    \tilde{A}_l^{(n)}&=\mathbb{P}(t\leq T)\e[\tilde{A}_l^{(n)}|t\leq T]+\mathbb{P}(t> T)\e[\tilde{A}_l^{(n)}|t> T],\\
     \tilde{B}_l^{(n)}&=\mathbb{P}(t\leq T)\e[\tilde{B}_l^{(n)}|t\leq T]+\mathbb{P}(t> T)\e[\tilde{B}_l^{(n)}|t> T].
\end{align}
Note that, given the reset event occurred before $T$ scheduling instances from the start of the sequence, we have the same age distribution at the beginning of $\tilde{A}_l^{(n)}$ and $\tilde{B}_l^{(n)}$. Hence, $\e[\tilde{A}_l^{(n)}|t\leq T]=\e[\tilde{B}_l^{(n)}|t\leq T]$. However, the age distributions will be different but with finite expectations, if the reset event occurs after $T$. In this case too, we can bound it from above using the age curve obtained for source-$n$ when the shared server is absent. Therefore, regardless of the initial distribution, we have
\begin{align}
\limsup_{L\to\infty}\frac{\e[\tilde{A}_l^{(n)}|t> T]}{(L-T)\e[Y]}\leq \tilde{\Delta}_n,\label{eqn:rst_A} \\
\limsup_{L\to\infty}\frac{\e[\tilde{B}_l^{(n)}|t> T]}{(L-T)\e[Y]}\leq \tilde{\Delta}_n. \label{eqn:rst_B}
\end{align}
Therefore, $\exists \ L_0$ such that for $L>L_0$, we have
\begin{align}
    \frac{\e[\tilde{A}_l^{(n)}|t> T]}{(L-T)\e[Y]}<2\tilde{\Delta}_n,\label{eqn:ineq_1}\\
    \frac{\e[\tilde{B}_l^{(n)}|t> T]}{(L-T)\e[Y]}<2\tilde{\Delta}_n.\label{eqn:ineq_2}
\end{align}
Then, from \eqref{eqn:rst_A}, \eqref{eqn:rst_B}, \eqref{eqn:ineq_1} and \eqref{eqn:ineq_2}, we have 
\begin{align}
    \left|\frac{\tilde{A}_l^{(n)}}{(L-T)\e[Y]}-\frac{\tilde{B}_l^{(n)}}{(L-T)\e[Y]} \right|<2\epsilon\tilde{\Delta}_n,
\end{align}
which yields
\begin{align}
    \left|\frac{A_l^{(n)}}{L\e[Y]}-\frac{B_l^{(n)}}{L{\e[Y]}}\right|<6\epsilon\tilde{\Delta}_n.
\end{align}
Now, let us denote the average age of the optimal PS as $\e[\Delta_{P}]$ and that of the optimal cyclic scheduler as $\e[\Delta_{C}]$. Then, 
\begin{align}
    \e[\Delta_{P}]&=\sum_nw_n\e[\Delta_n],\\
    &=\sum_nw_n\sum_l\beta_l\frac{A_l^{(n)}}{L\e[Y]},\\
    &\geq \sum_n\sum_lw_n\beta_l\left(\frac{B_l^{(n)}}{L\e[Y]}-6\epsilon\tilde{\Delta}_n\right),\\
    &=\sum_l\beta_l\sum_nw_n\frac{B_l^{(n)}}{L\e[Y]}-6\epsilon\sum_l\sum_n\beta_lw_n\tilde{\Delta}_n,\\
    &\geq\min_l \sum_nw_n\frac{B_l^{(n)}}{L\e[Y]} -6\epsilon\sum_l\sum_n\beta_lw_n\tilde{\Delta}_n,\\
    &\geq\e[\Delta_C]-6\epsilon\sum_l\sum_n\beta_lw_n\tilde{\Delta}_n.
\end{align}
Therefore, we have that for all $\epsilon>0$, $\e[\Delta_{P}]+\epsilon\geq\e[\Delta_C]$ proving the desired result.
\color{black}

\section*{Acknowledgment}
This work is done when N.~Akar is on sabbatical leave as a visiting professor at University of Maryland, MD, USA, which is supported in part by the Scientific and Technological Research Council of T\"{u}rkiye  (T\"{u}bitak) 2219-International Postdoctoral Research Fellowship Program.

\bibliographystyle{unsrt} 
\bibliography{refs}

\end{document}